\documentclass[a4paper,fleqn]{cas-dc}

\usepackage{lineno}
\modulolinenumbers[1]
\usepackage[inline]{trackchanges} 

\usepackage[authoryear]{natbib}

\usepackage{upgreek}
\newcommand{\mum}{$\upmu$m}

\graphicspath{{Figures/}}

\newcommand{\reff}{$r_\text{eff}$}
\newcommand{\kext}{$k_\text{ext}$}
\newcommand{\taucld}{$\tau_\text{cld}$}
\newcommand{\taucldzero}{$\tau_\text{cld, 0}$}
\newcommand{\taucldacs}{$\tau_\text{cld, ACS}$}

\def\tsc#1{\csdef{#1}{\textsc{\lowercase{#1}}\xspace}}
\tsc{WGM}
\tsc{QE}

\begin{document}
\let\WriteBookmarks\relax
\def\floatpagepagefraction{1}
\def\textpagefraction{.001}

\shorttitle{Simultaneous clouds observations from EMM/EXI and TGO/ACS-MIR}    

\shortauthors{Stcherbinine et al.}  

\title [mode = title]{On the impact of the vertical structure of Martian water ice clouds 
on nadir atmospheric retrievals from simultaneous EMM/EXI and TGO/ACS-MIR observations.}



%

\author[1,2]{Aurélien Stcherbinine}
    [style=french,
    orcid=0000-0002-7086-5443]

\cormark[1]


\ead{Aurelien.Stcherbinine@irap.omp.eu}

\ead[url]{https://aurelien.stcherbinine.net}


\author[3]{Michael J. Wolff}
    [style=english,
    orcid=0000-0002-1127-8329]

\author[1]{Christopher S. Edwards}
    [style=english,
    orcid=0000-0002-8096-9633]

\author[4]{Oleg Korablev}
    [style=english,
    orcid=0000-0003-1115-0656]

\author[4]{Anna Fedorova}
    [style=english,
    orcid=0000-0002-4176-2955]

\author[4]{Alexander Trokhimovskiy}
    [style=english,
    orcid=0000-0003-4041-4972]
    
\affiliation[1]{organization={Department of Astronomy and Planetary Science, Northern Arizona University},
            city={Flagstaff},
            state={AZ},
            country={USA}}

\affiliation[2]{organization={Institut de Recherche en Astrophysique et Planétologie, CNES, Université Toulouse III Paul Sabatier, CNRS},
            city={Toulouse},
            country={France}}

\affiliation[3]{organization={Space Science Institute},
            city={Boulder},
            state={CO},
            country={USA}}

\affiliation[4]{organization={Space Research Institute (IKI)},
            city={Moscow},
            country={Russia}}

\cortext[1]{Corresponding author}



\begin{abstract}
Retrieving the optical depth of the Martian clouds (\taucld) is a powerful way to monitor their spatial and
temporal evolution. However, such retrievals from nadir imagery rely on several assumptions,
including the vertical structure of the clouds in the atmosphere.
Here we compare the results of cloud optical depth retrievals at 320~nm from the Emirates eXploration Imager (EXI)
onboard the Emirates Mars Mission (EMM) "Hope" orbiter performed using a basic uniform cloud profile used in previous studies and using derived cloud profiles obtained from near-simultaneous Solar
Occultation observations in the 3.1--3.4~\mum{} spectral range from the Middle-Infrared channel of the Atmospheric Chemistry Suite (ACS)
instrument onboard the ESA Trace Gas Orbiter (TGO).
We show that the latitudinal dependence of the cloud vertical profiles can have a strong impact on the nadir retrievals; neglecting it can lead to
a significant underestimation of \taucld{} in the polar regions
(up to 25~\% to 50~\%, depending on the vertical distribution of the dust in the atmosphere)
and to a lesser extent, to an overestimation of \taucld{} around the equator.
We also discuss the impact of a vertically-dependent particle size profile,
as previous studies have shown the presence of very small water ice particles at the top of the clouds.
From this analysis, we provide recommendations for the improvement of water ice cloud parameterization in radiative transfer algorithms in nadir atmospheric retrievals.

\end{abstract}




\maketitle

\section{Introduction}  \label{sec:introduction}

    Water ice clouds play an important role in the current Martian climate: they affect the thermal structure
    of the atmosphere by absorbing and scattering the incoming Solar radiation \citep{haberle_2011, madeleine_2012, 
    navarro_2014a, wilson_2007, wilson_2008}, and are a major actor in the inter-hemispheric water exchange
    with a different dynamic compared to that of water vapor \citep{clancy_1996, montmessin_2004, montmessin_2017}.
    Constraining the properties and behavior of the water ice clouds on Mars is thus of importance to a better
    understanding of the current climate of the planet and its evolution.

    Over the past 25~years, several orbital missions and their instruments have been looking at the Martian clouds through
    different geometries: nadir in UV-Visible \citep[Mars Color Imager (MARCI), Emirates eXploration Imager (EXI); e.g.,][]{wolff_2019, wolff_2022}, thermal 
    \citep[Thermal Emission Spectrometer (TES), Thermal Emission Imaging System (THEMIS), Mars Climate Sounder (MCS), Emirates Mars Infrared Spectrometer (EMIRS); e.g.,][]{smith_2001a, smith_2003, smith_2022, heavens_2011a, atwood_2022} or infrared wavelength 
    \citep[Observatoire pour la Minéralogie, l'Eau, les Glaces et l'Activité (OMEGA), Compact Reconnaissance Imaging Spectrometer for Mars (CRISM), Spectroscopy for Investigation of Characteristics of the Atmosphere of Mars (SPICAM); e.g.,][]{vincendon_2011, szantai_2021, mateshvili_2007, willame_2017},
    limb \citep[OMEGA, CRISM, SPICAM, Imaging Ultraviolet Spectrograph (IUVS); e.g.,][]{vincendon_2011a, clancy_2019, rannou_2006, stevens_2017}, or Solar Occultation 
    \citep[SPICAM, Atmospheric Chemistry Suite (ACS), Nadir and Occultation for MArs Discovery (NOMAD); e.g.,][]{montmessin_2017a, maattanen_2013, stcherbinine_2020, stcherbinine_2022e, luginin_2020, liuzzi_2020, stolzenbach_2023a}.
    As several missions remain active at present, one can now make direct comparisons of their respective datasets, with information from their different viewing geometries complementing each other. This results in a more comprehensive view of Martian clouds.

    Nadir observations with UV-visible multi-band cameras such as MARCI onboard the
    Mars Reconnaissance Orbiter (MRO) \citep{malin_2001, belliii_2009} or EXI
    of the Hope probe \citep{jones_2021, amiri_2022} provide a large spatial coverage per image but only
    limited access to the vertical structure of the aerosol profile in the atmosphere. 
    Similarly, ground-based observations from Earth also provide a global view of the spatial distribution of Martian clouds, but with a lower resolution and again very limiter access the vertical structure of the atmosphere \citep{lilensten_2022}.
    This is in contrast to Solar Occultation (hereafter "SO") or limb-viewing geometry
    for instance, that provide detailed vertical profiles but only with a limited spatial extent (especially in the case of SO).
    Thus, nadir imagery is the easier way to constrain the spatial distribution and variations of the Martian clouds.
    However, retrieving the column-integrated optical depth of the clouds (\taucld) from nadir reflectance measurements
    requires assumptions on the vertical structure and properties of the atmosphere, including the vertical 
    distribution of dust and ice aerosols.
    Such assumptions on the aerosol properties can be a significant source of retrieval uncertainties \citep[e.g.,][]{wolff_2019}.

    Typically, the retrieval algorithms for MARCI and EXI assume clouds characterized by water ice particles of the same size uniformly distributed
    from a given bottom altitude (usually taken between 15 and 20~km) to 100~km, the bottom altitude being the same regardless of the time of the year or the
    geographic coordinates of the clouds \citep[e.g.,][]{wolff_2019, wolff_2022}.
    An alternative approach that has been implemented for THEMIS and TES cloud retrievals implemented a
    variable lower altitude of the clouds, using the water condensation altitude as the bottom
    boundary for the presence of water ice \citep[e.g.,][]{clancy_1996, smith_2003, wolff_2003}.
    However, recent studies have shown that the cloud altitude varies significantly with the season and latitude
    \citep[e.g.,][]{stcherbinine_2022e, smith_2013, montmessin_2006, heavens_2011a, lolachi_2022},
    and that the distribution of the ice particle sizes with the altitude is characterized typically by a decrease in size when the altitude increases \citep[e.g.,][]{clancy_2019, 
    stcherbinine_2020, stcherbinine_2022e, luginin_2020, liuzzi_2020, stolzenbach_2023a}.
    Indeed, if water ice clouds are observed mostly from 10 to 40~km in the polar regions around $L_s~\sim~180^\circ$, they are detected between 40 and 80~km at the same time around the equator, and between 20 and 50~km in the equatorial regions around $L_s~\sim~90^\circ$ \citep{stcherbinine_2022e}.
    This observed latitudinal and temporal diversity in the altitude of the water ice particles is in
    contrast with the classical approach of using always the same single vertical profile for all the clouds.
    Thus, the question that arises is: how do more realistic vertical profiles of water ice cloud properties (i.e.,
    obtained with SO geometry) affect the optical depth retrievals from nadir data compared to those using more simple vertical prescriptions?

    In this article, we combine simultaneous ACS-MIR vertical profiles of Martian water ice clouds and nadir images acquired by
    EXI to study the impact of refining the vertical structure of the clouds on the EXI optical depth retrievals.
    First, we present in Section~\ref{sec:methods} the datasets and methods used in this study. 
    Then, Section~\ref{sec:discussion} discusses the results on the impact of altitude variations and particle size gradient in
    the EXI retrievals.
    Finally, Section~\ref{sec:ccl} summarizes the main points of the study.

\section{Data and Methods}  \label{sec:methods}

    \subsection{Vertical clouds profiles from ACS-MIR}  \label{sec:methods_acs}
        The Atmospheric Chemistry Suite (ACS) instrument is a set of three spectrometers onboard the ExoMars Trace Gas Orbiter (TGO) ESA-Roscosmos spacecraft, which has been conducting science operations for 3 Martian Years (MY), since March 2018 \citep{korablev_2018, korablev_2019, vandaele_2019}.
        The Mid-InfraRed (MIR) channel is a high-resolution cross-dispersion echelle spectrometer dedicated to SO geometry
        \citep{trokhimovskiy_2015a, korablev_2018}.

\begin{figure}[h]
            \centering
            \includegraphics[width=\columnwidth]{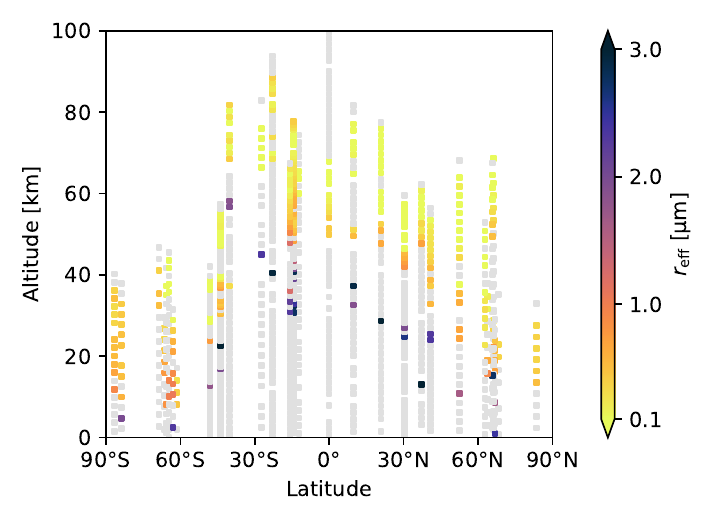}
            \caption{The vertical profiles of water ice clouds effective radius (\reff) as a
                function of altitude derived from ACS-MIR measurements
                used in this study, as a function of the latitude of the observation
                (see Figure~\ref{fig:acs_lon_lat_map} for the spatial distribution of the 
                observations).
                $L_s$ are ranging from 76.4$^\circ$ to 358.8$^\circ$ over MY~36
                (see details of the observations in Table~\ref{tab:obs_list}).
                Data from the first half of MY~36 have already been published in
                \citet{stcherbinine_2022_dataset}.
                Observations without water ice detections are in gray.}
            \label{fig:acs_profiles}
        \end{figure}

\begin{figure*}[h]
            \centering
            \includegraphics[width=\textwidth]{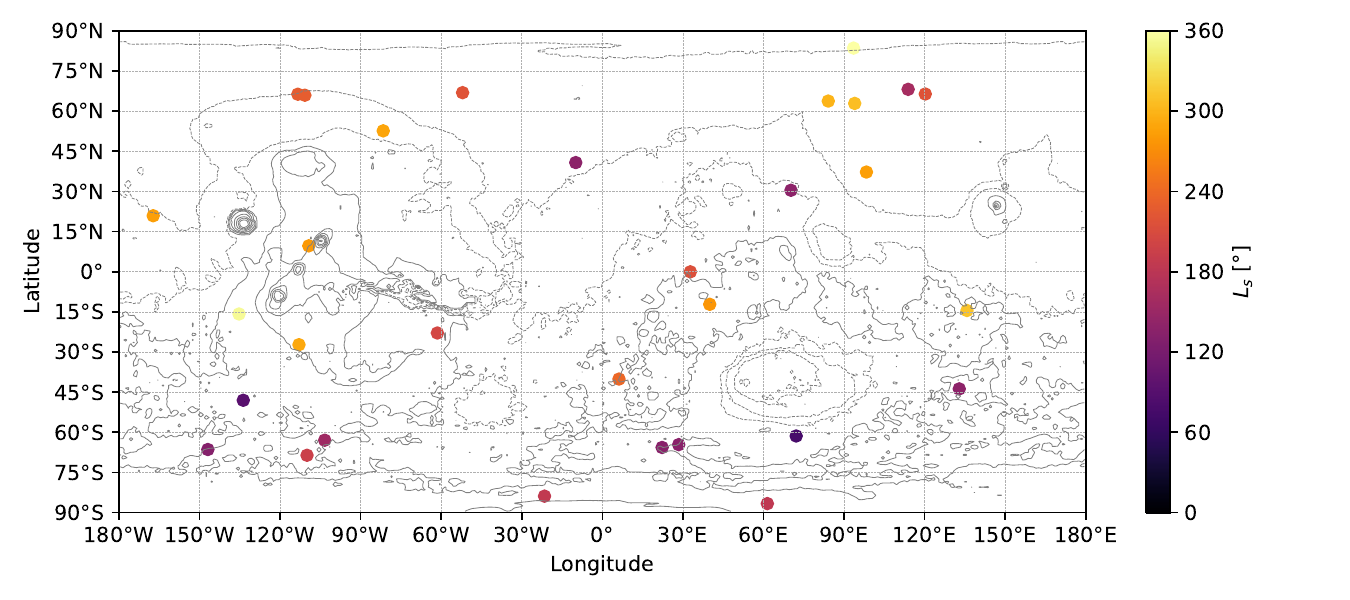}
            \caption{Spatial (longitude and latitude) and temporal ($L_s$, spanning over MY~36) distribution of the 
                31 ACS-MIR observations used in this study. See Table~\ref{tab:obs_list} for more details.}
            \label{fig:acs_lon_lat_map}
        \end{figure*}

        We use the methodology and results presented in \citet{stcherbinine_2020, stcherbinine_2022e} and summarized in the following, to retrieve the water ice opacity, as well as the ice particle effective size, as a function of altitude in the Martian atmosphere. The observations use the 3-\mum{} region observed with the ACS-MIR position 12 of the primary grating mirror (covering wavelengths from 3.1 to 3.4~\mum{}) and have a vertical resolution of $\sim$2.5~km.
        The atmospheric transmittance values measured by the instrument at each observed altitude (tangent point) are
        converted into extinction coefficients (\kext) through an onion-peeling (vertical)
        inversion algorithm \citep{goldman_1979}. The resulting extinction spectra are fitted 
        with models using spherical water ice and dust particles of various sizes in order to
        constrain the presence and the size of water ice particles in the atmosphere.  
        Some ACS clouds profiles, from the first half of MY~36, have already been published in 
        \citet{stcherbinine_2022_dataset}, but more recent observations have been
        processed specifically for this study.
        The ACS-MIR clouds profiles used in this study are presented as a function of the latitude
        in Figure~\ref{fig:acs_profiles}, and Figure~\ref{fig:acs_lon_lat_map}
        shows the spatial and temporal distribution of these observations.
        
        An unfortunate limitation of the method, due to the small wavelength range covered by the position 12 acquisitions (3.1 -- 3.4~\mum), is that it is not sensitive to the larger water ice particles (\reff{} $\gtrsim$ 3~\mum) that
        may be present at the bottom of the profiles; that is to say that one will not be able to distinguish them 
        from dust particles (see discussion in \citet{stcherbinine_2020, stcherbinine_2022e}).
        However, these potential layers at low altitudes will likely be mixed with a larger amount of dust \citep[e.g.,][]{smith_2013}, 
        which will limit their contribution to the clouds' optical depth as observed from the orbit (see discussion in
        Section~\ref{sec:discussion}).

    \subsection{Optical depths retrievals with EXI}     \label{sec:methods_exi}
        The Emirates eXploration Imager (EXI) instrument onboard the Emirates Mars Mission (EMM)
        "Hope" probe is a UV-Visible framing camera that has been observing the full Martian disk in 
        6 bandpasses since the beginning of its science phase in May 2021 
        \citep{amiri_2022, jones_2021}.
        Here, we follow the method of \citet{wolff_2022} that uses the normalized reflectance (I/F) at 
        $\lambda=320$~nm (the EXI f320 band) to retrieve the water ice clouds
        optical depth \taucld.
        The EXI data used in this study have been acquired in the so-called XOS-1 mode with
        a $2\times2$ binning, giving a spatial resolution of 4--8~km per pixel at nadir, and
        processed to level 2A. Information about the EXI data products is provided in 
        the Data Availability section.
        
        We use the DIScrete Ordinate Radiative Transfer (DISORT) code 
        \citep{stamnes_1988, stamnes_2017} through the
        \emph{pyRT\_DISORT} Python module \citep{connour_2023} to perform the retrievals at 320~nm.
        In terms of the model parameters, we employ the following, based on \citet{wolff_2019, wolff_2022}:
        \begin{itemize}
            \item Surface: Hapke reflectance model, with the values of the parameters 
                of the MARCI band 7 model from \citet{wolff_2019},
                scaled to be adjusted to the EXI radiometric calibration \citep{wolff_2022}.
            \item Atmospheric dust: 
                a vertical profile with a constant volume mixing ratio (with respect to the CO$_2$
                gas density) of 1.5~\mum{} dust particles from the surface to 100~km as it is usually the
                case in such studies \citep[e.g.,][]{wolff_2017, wolff_2019, wolff_2022}
                (or alternative upper altitude derived from ACS-MIR observations).
                Dust properties are computed assuming asymmetric hexahedral shapes with a sphericity of 0.78
                and an effective variance $\nu_\text{eff}$ of 0.3 \citep{saito_2021}, and refractive indices
                from \citet{wolff_2010, connour_2022a}.
                The total column optical depth 
                $\tau_\text{dust}$ is extracted from a contemporaneous dust climatology constructed
                from the EMIRS observations \citep{smith_2022, edwards_2021}.
            \item Pseudo-spherical atmosphere geometry: as ACS-MIR profiles are acquired close to the terminator
                (see section~\ref{sec:methods_cross_obs}), they are associated with high incidence
                and potentially with large emergence angles as well. We use the pseudo-spherical correction
                introduced in DISORT 3 \citep{stamnes_2017} instead of the classical plane-parallel representation, allowing for an explicit treatment of planetary/atmospheric curvature.
            \item Water ice clouds: extinction and scattering properties of water ice particles are
                computed assuming droxtals shapes, a gamma size distribution \citep{hansen_1974}
                with an effective variance of 0.1 \citep[e.g.,][and reference contained within]{wolff_2019}.
        \end{itemize}
        
        Typically, these kinds of retrievals assume a vertical profile of the water ice aerosols 
        defined by a cloud bottom around 20~km and a constant volume mixing ratio above this altitude 
        \cite[e.g.,][]{wolff_2019, wolff_2022}.
        In the present study, we will compare the optical depth values retrieved using this
        cloud vertical profile on some examples, with results using alternative vertical structures 
        (altitude, ice particle sizes) from ACS-MIR coincident observations.

    \subsection{Cross-observations}     \label{sec:methods_cross_obs}
        While EXI offers column-integrated cloud optical depths over the
        illuminated Martian disk, ACS-MIR will provide locally detailed information about the
        vertical distribution of the water ice particle properties (cf Figure~\ref{fig:acs_profiles}).
        Due to the SO observing geometry, ACS-MIR can only acquire atmospheric profiles near
        the morning and evening terminators (local times $\sim$ 06:00 and 18:00).

        To perform optical depth retrievals in nadir geometry, the vertical distribution of the
        aerosols in the atmosphere is one of the main assumptions that must be made.
        Usually, one uses a uniform distribution of one size distribution with the base of the cloud 
        starting always at the same given altitude regardless of the latitude
        or season (typically taken between 15 and 20~km) \citep[e.g.,][]{wolff_2017, wolff_2019, wolff_2022}.

        Even though no coordinated observations between the EXI and ACS
        instruments (or the EMM and TGO probes) have been explicitly planned, they are both looking at the
        Martian atmosphere from their respective orbits. 
        Thus, we searched for observations from both instruments that observed the 
        same coordinates within one hour between February 2021 and January 2023, which includes the
        whole MY 36. We identify 31 observations from each instrument that match these criteria,
        spanning from $L_s=76^\circ$ to $L_s=359^\circ$ (MY 36) and covering all latitudes
        from $86^\circ$S to $86^\circ$N.
        Table~\ref{tab:obs_list} gives the list and coordinates of the observations used in this study
        and Figure~\ref{fig:acs_lon_lat_map} shows the spatial distribution
        of the ACS-MIR observations on a map of Mars.

        \begin{table*}[width=.9\textwidth, cols=6, pos=h]
            \caption{List and coordinates of the 31 EXI and ACS-MIR observations used in this
                study. All observations have been acquired over MY~36.}
            \label{tab:obs_list}
            \begin{tabular*}{\tblwidth}{@{}LLRRR@{}}
                \toprule
                EXI (XOS-1 -- f320) & ACS-MIR & Longitude [°E] & Latitude [°N] & $L_s$ [°]  \\ 
                \midrule
                20210724T220055 & ORB016367\_N1 &   72.2 & -61.4 &  76.4 \\ 
                20210830T063840 & ORB016809\_N2 & -133.7 & -48.0 &  92.3 \\
                20211124T202544 & ORB017868\_N1 & -146.9 & -66.4 & 132.2 \\
                20211129T120020 & ORB017925\_N1 &   22.2 & -65.7 & 134.4 \\
                20211201T132848 & ORB017950\_N1 &   28.4 & -64.6 & 135.4 \\
                20211204T223121 & ORB017991\_N2 &   -9.9 &  40.8 & 137.1 \\
                20211206T172533 & ORB018013\_N2 &   70.2 &  30.4 & 138.0 \\
                20211208T130036 & ORB018035\_N1 &  132.9 & -43.8 & 138.8 \\
                20220107T191342 & ORB018403\_N1 & -103.4 & -63.0 & 154.1 \\
                20220121T212521 & ORB018576\_N1 &  113.9 &  68.1 & 161.6 \\
                20220303T020711 & ORB019068\_N1 &   61.4 & -86.6 & 183.7 \\
                20220307T100322 & ORB019121\_N1 &  -21.5 & -83.9 & 186.2 \\
                20220323T024006 & ORB019312\_N1 & -110.0 & -68.6 & 195.3 \\
                20220407T071615 & ORB019498\_N1 &  -61.4 & -22.9 & 204.3 \\
                20220424T012729 & ORB019701\_N2 &   32.8 &  -0.1 & 214.5 \\
                20220429T075125 & ORB019766\_N1 &  120.2 &  66.4 & 217.7 \\
                20220501T204754 & ORB019797\_N1 &  -52.0 &  66.9 & 219.3 \\
                20220515T082017 & ORB019962\_N1 & -113.4 &  66.3 & 227.7 \\
                20220517T101316 & ORB019987\_N1 & -110.8 &  65.9 & 229.0 \\
                20220531T055303 & ORB020155\_N2 &    6.2 & -40.1 & 237.7 \\
                20220801T045738 & ORB020913\_N1 &   40.0 & -12.2 & 276.9 \\
                20220805T165821 & ORB020968\_N1 & -109.3 &   9.7 & 279.7 \\
                20220811T143523 & ORB021039\_N2 & -167.2 &  20.9 & 283.3 \\
                20220812T064308 & ORB021048\_N1 &   98.3 &  37.2 & 283.8 \\
                20220818T212312 & ORB021129\_N1 &  -81.6 &  52.6 & 287.8 \\
                20220822T161908 & ORB021174\_N1 & -113.0 & -27.3 & 290.1 \\
                20220906T211836 & ORB021361\_N1 &   84.1 &  63.7 & 299.4 \\
                20220918T043109 & ORB021499\_N1 &   94.0 &  62.9 & 306.1 \\
                20220925T092024 & ORB021587\_N1 &  135.8 & -14.5 & 310.3 \\
                20221213T193446 & ORB022556\_N2 & -135.3 & -15.8 & 353.6 \\
                20221223T220315 & ORB022680\_N1 &   93.5 &  83.5 & 358.8 \\
                \bottomrule
            \end{tabular*}
        \end{table*}

\section{Results and Discussion}    \label{sec:discussion}

    The retrievals of water ice cloud optical depth are performed on the 31 EXI pixels associated
    with an ACS-MIR profile (cf. Table~\ref{tab:obs_list}).
    For all of them, we run the retrieval algorithm assuming the same typical uniform cloud profile
    (\reff~=~3~\mum{} between 20 and 100~km) and profiles adjusted with inputs from ACS-MIR observations
    (altitude, size).
    In Section~\ref{sec:impact_altitude} we discuss the impact of the altitude of the clouds on the retrievals, and in Section~\ref{sec:impact_size} we also introduce
    a size gradient for the ice within the clouds.

    Regarding the dust, we run the retrievals with uniform profiles of 1.5~\mum{} particles uniformly
    distributed (with respect to the gas density) from the surface to two alternative altitudes for all of the observations:
    both 100~km as it is usually the case in such studies \citep[e.g.,][]{wolff_2017, wolff_2019, wolff_2022}
    and, when using cloud altitudes from ACS-MIR profiles, the highest altitude without water ice detection
    below the clouds in the ACS-MIR profiles. Indeed, previous studies have shown that the dust in the
    atmosphere is essentially present below the higher water ice clouds \citep{smith_2013, luginin_2020, liuzzi_2020, stolzenbach_2023a}. This is particularly important to consider for low-altitude clouds (in the polar regions)
    because the higher amount of dust particles at low altitudes will have a stronger impact on the \taucld{}
    retrievals: more dust at the cloud altitude or above will "hide" the water ice by absorbing and scattering
    the light scattered by the water ice clouds, resulting in higher \taucld{} in the retrievals.

    For simplification purposes, in the following, even though some ACS-MIR profiles can exhibit multiple cloud layers 
    \citep{stcherbinine_2020, stcherbinine_2022e}, we will consider only one continuous cloud layer
    per profile for the retrievals, with the bottom and top altitudes corresponding respectively to
    the lowest and highest ice detections in the ACS-MIR atmospheric profile.
    However, if a profile exhibits multiple layers of water ice, we will consider that the maximum
    altitude for the dust is below the upper layer. Below this altitude, we will have a mix of dust 
    and water ice particles, which is representative of what can be observed by the orbiters 
    \citep[e.g.,][]{smith_2013, stcherbinine_2020, luginin_2020, stolzenbach_2023a}.

    \subsection{Cloud altitudes}    \label{sec:impact_altitude}
        \begin{figure*}[h]
            \centering
            \includegraphics[width=\textwidth]{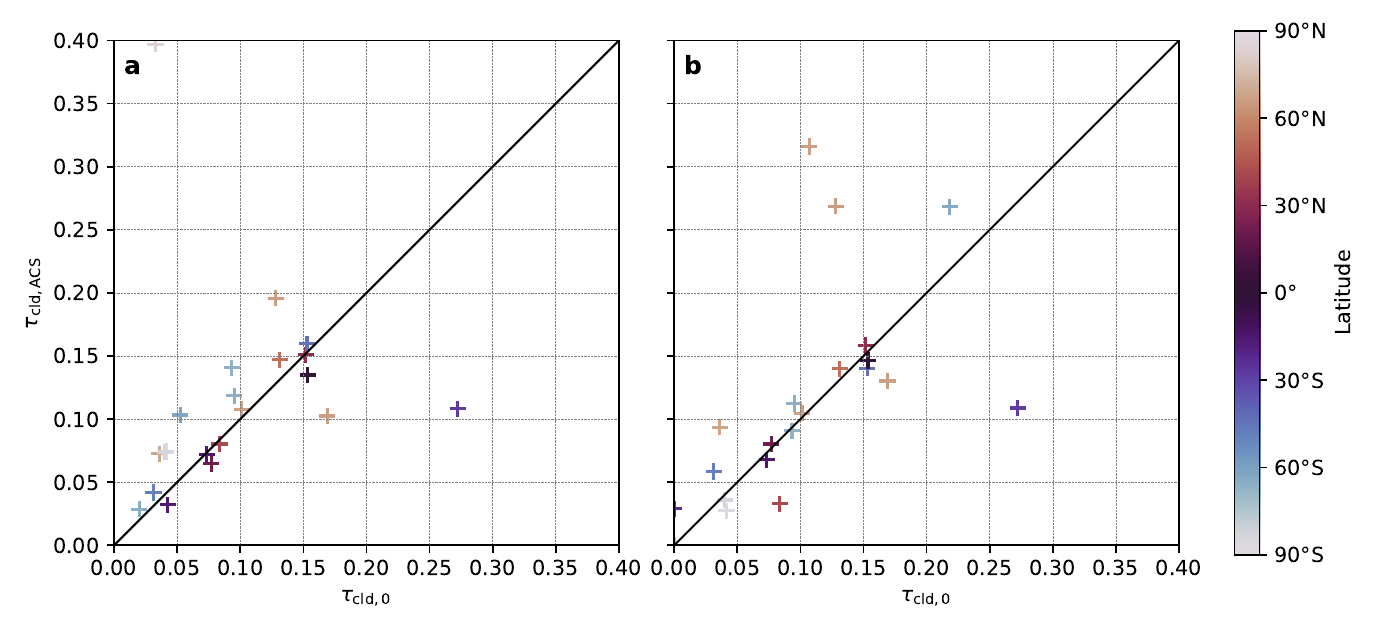}
            \caption{EXI optical depth retrieved using clouds altitudes from ACS-MIR 
                observations ($\tau_\text{cld, ACS}$) as a function of the optical depth retrieved
                using the typical cloud profile from 20 to 100~km ($\tau_\text{cld, 0}$).
                The color of the crosses shows the latitude of the associated observation,
                and the black line represents the 
                $\tau_\text{cld, ACS} = \tau_\text{cld, 0}$ line.
                \textbf{a.} Dust particles are uniformly distributed from 0 to 100~km.
                \textbf{b.} Dust is only located below the clouds when using ACS-MIR altitudes.}
            \label{fig:tau_acs_vs_0}
        \end{figure*}

        \begin{figure*}
            \centering
            \includegraphics[width=\textwidth]{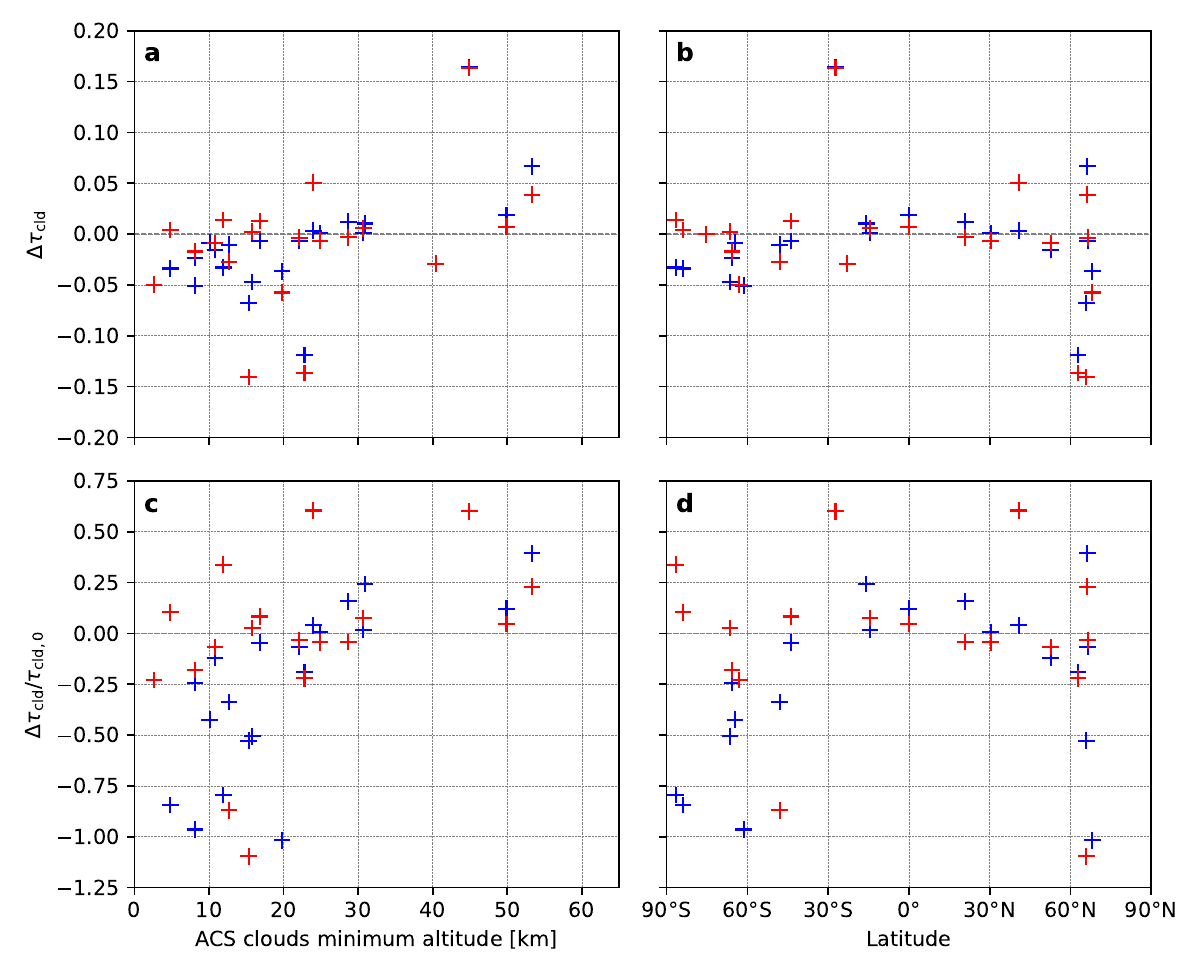}
            \caption{\textbf{a. \& b.}
                Difference between the \taucld{} values retrieved from EXI observations 
                assuming water ice clouds with particle size of 3~\mum{} from 20~km to 100~km
                (\taucldzero)
                and the ones retrieved using top and bottom altitudes from ACS-MIR profiles 
                (\taucldacs)
                and dust with particle size of 1.5~\mum{} from 0~km to 100~km (\textcolor{blue}{blue}),
                or dust only contained below the clouds (\textcolor{red}{red}).
                $\Delta\tau_\text{cld} = \tau_\text{cld, 0} - \tau_\text{cld, ACS}$
                \textbf{c. \& d.} Same but normalized by the \taucld{} values retrieved with
                clouds uniform between 20~km and 100~km (i.e., no ACS input).
                Results are displayed as a function of the clouds' minimum altitude from the
                ACS-MIR profiles (a \& c) and the latitude of the observation (b \& d).
                }
            \label{fig:comp_taucld_alt_acs}
        \end{figure*}

        We define the following notations:
        \begin{itemize}
            \item $\tau_\text{cld, 0}$: optical depth retrieved assuming a typical constant profile with 3~\mum{}
                water ice crystals uniformly distributed from 20~km to 100~km.
            \item $\tau_\text{cld, ACS}$: optical depth retrieved using clouds altitudes taken from ACS observations.
            \item $\Delta\tau_\text{cld} = \tau_\text{cld, 0} - \tau_\text{cld, ACS}$
        \end{itemize} 

        Figure~\ref{fig:tau_acs_vs_0} shows the EXI optical depth retrieved
        assuming 3~\mum{} water ice particles at altitudes taken from ACS-MIR observations
        ($\tau_\text{cld, ACS}$) as a function of the optical depth retrieved using the typical
        constant profile from 20 to 100~km ($\tau_\text{cld, 0}$).
        We can see that globally, points associated with near-equatorial latitudes (darker points) are closer
        to the $y=x$ line and slightly below ($\tau_\text{cld, ACS} < \tau_\text{cld, 0}$) 
        compared to the more polar observations (lighter points) that are more clearly above this line
        ($\tau_\text{cld, ACS} > \tau_\text{cld, 0}$).
        Adjusting the maximum altitude of the dust particles according to the altitude of the clouds
        for $\tau_\text{cld, ACS}$ (panel b) has two main effects on the distribution of the results:
        less scattering of around the $y=x$ line for retrievals associated with $\tau \sim 0.9-0.18$,
        but more difference between the two cases $\tau_\text{cld, ACS}$ and $\tau_\text{cld, 0}$ for
        smaller or larger values, especially for retrievals associated with observations performed with
        more polar latitudes.

        Figure~\ref{fig:comp_taucld_alt_acs} shows the difference ($\Delta$\taucld, panels a \& b) and relative
        difference ($\Delta$\taucld / \taucldzero, panels c \& d) between EXI water ice clouds retrievals performed with
        the same typical cloud profile of 3~\mum{} water ice particles uniformly distributed between
        20 and 100~km, and with profiles whose altitudes are taken from contemporaneous
        ACS-MIR observations.
        Results are presented as a function of the minimum cloud altitude derived from the ACS-MIR
        observations (panels a \& c) and the latitude of the observation (panels b \& d).
        For the blue crosses, the bottom and top altitudes of the clouds are taken from ACS-MIR
        data but the upper altitude for the dust is still 100~km, and for the red crosses 
        the dust is only present below the water ice clouds (with the minimum and maximum
        altitudes from ACS-MIR data).
        One can note that the total column-integrated dust optical depth remains the same
        in both cases, as it is taken from contemporaneous EMIRS observations.

        We can see that the retrieved value of \taucld{} depends on the altitude of the cloud
        provided in the model: for a given set of atmospheric and surface parameters, 
        $\Delta$\taucld{} decreases and increases along with the minimum altitude of the clouds, i.e.,
        the value of
        \taucld{} increases when the minimum altitude of the cloud decreases.
        Thus, if not accounting for the altitude of the clouds, the retrieval algorithm will underestimate the optical depth of the low-altitude clouds
        ($\Delta\tau_\text{cld} < 0$)
        and overestimate it for the high-altitude clouds 
        ($\Delta\tau_\text{cld} > 0$), relatively to the bottom altitude assumed in 
        the model, i.e., $\sim$~20~km.
        For clouds starting below $\sim$~10~km, \taucld{} will be underestimated by 25~\%
        (up to 0.05 in absolute value), and similarly, for clouds with a bottom altitude between
        40 and 60~km the \taucld{} will be overestimated up to 50~\%
        (cf. Figure~\ref{fig:comp_taucld_alt_acs} a \& c).
        We also observe on Figure~\ref{fig:comp_taucld_alt_acs} that, as expected, the retrievals
        are more affected by the dust profile for clouds with a minimum altitude below $\sim$~20~km
        (more difference between the blue and red points),
        and that the retrieved values of \taucld{} decrease (i.e., $\Delta\tau_\text{cld}$ increase) when the dust is capped by the water ice clouds (red points are generally above the blue ones).

        The altitude of the clouds varies significantly with the latitude
        \citep{forget_1999, heavens_2011a, jaquin_1986, montmessin_2006, smith_2013,
        stcherbinine_2022e, lolachi_2022}, which is visible in Figure~\ref{fig:acs_profiles}.
        \citet{stcherbinine_2022e} reported that the typical altitude of the clouds
        is 20--40~km higher around the equator than in the polar regions, and that variations
        of the same order of magnitude also occur between summer and winter for both hemispheres.
        In our dataset, the minimum altitude of the clouds ranges from 3~km to 53~km.
        However, we do not have enough data here to discuss the impact of seasonal variations
        of the clouds.

        As a consequence of the latitudinal variations of the cloud altitudes, the EXI nadir
        retrievals will tend to underestimate the \taucld{} in the polar regions (up to $\sim$~25~\%
        if also decreasing the maximum altitude of the dust; $\sim$~50~\% otherwise)
        and slightly overestimate it around the equator (up to $\sim$~10~\% if also decreasing the 
        maximum altitude of the dust; $\sim$~25~\% otherwise)
        if assuming the same uniform vertical profile between 20 and 100~km for all clouds 
        (cf. Figure~\ref{fig:comp_taucld_alt_acs} b \& d).
        We can see that the red points have a smaller vertical scatter between the poles and the equator
        compared to the blue ones, which means that
        the amplitude of the variations in \taucld{} when adjusting the
        altitude of the clouds is lowered if keeping the dust particles below the clouds.
        One should also note that the largest relative discrepancies between the retrievals observed in the
        polar regions are associated with absolute differences in the \taucld{} values smaller than 0.05.
        The presence of optically thinner clouds in these regions contributes to enlarging the relative
        differences between the retrievals with and without taking into account the clouds and dust altitudes,
        but even the absolute variations $\Delta$\taucld{} are still higher under polar latitudes as
        shown on Figure~\ref{fig:comp_taucld_alt_acs}b.

        The sensitivity of \taucld{} with the cloud altitudes is related to the presence of dust
        in the atmosphere. At low altitudes, the higher concentration of dust particles in the atmosphere 
        will hide a larger fraction of the light scattered by the clouds. Thus, more water ice
        is required to reproduce the radiance observed from the orbit, which leads to thicker clouds
        in the retrievals. This explains the higher \taucld{} values retrieved close to the poles when
        using the ACS altitudes for the clouds in the retrieval algorithm, where they are observed down
        to altitudes lower than 20~km.
        Also, as the density of dust particles decreases with the altitude (with respect to the CO$_2$ gas 
        density) the distribution of the ice particles at high altitudes (typically above 50--60~km)
        will have a more limited impact on the \taucld{} retrievals as these optically thinner layers will
        have a more minor contribution in the total column-integrated optical depth value.
        This explains why we observe such large variations in the retrieved \taucld{} in the polar
        regions between computations with the dust altitude ranging up to 100~km, or only below the clouds.
        For instance, if we look at the two clouds 87$^\circ$S and 84$^\circ$S, their ACS-MIR profiles
        report water ice between $\sim$~10~km and $\sim$~35~km, with the upper water ice layer starting
        from $\sim$~15~km. So at these altitudes where the particle density is higher (for both ice and dust), a few kilometers
        of dust mixed with water ice will have a strong impact on the retrievals by hiding the light
        reflected by the water ice particles.

    \subsection{Clouds particle sizes}   \label{sec:impact_size}
        We discussed in the previous section the impact of the altitude of the clouds on the EXI retrievals for a
        fixed value of \reff. Here we use the altitude of the clouds as observed by ACS-MIR to focus on the impact
        of the size of the ice particles on the retrievals.
        More specifically, we focus on the vertical variations of the particle size within the cloud, rather
        than the choice of 3~\mum{} as the default reference size for \reff{}.
        Indeed, observational profiles of the Martian clouds by ACS-MIR have shown that they exhibit a decrease
        of the \reff{} when the altitude increases, with sizes typically ranging from 2-3~\mum{} at the bottom to
        0.1~\mum{} at the top \citep[Figure 4]{stcherbinine_2022e}.

        \begin{figure}[h]
            \centering
            \includegraphics[width=\columnwidth]{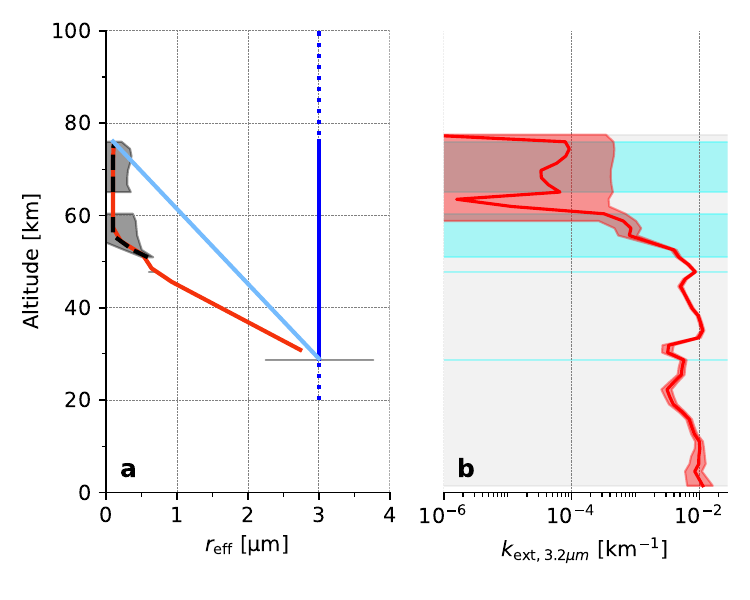}
            \caption{\textbf{a.} Vertical \reff{} profile (black dashed line) of a Martian cloud derived from ACS-MIR observation ORB021039\_N2
            (see Table~\ref{tab:obs_list}) with the associated uncertainties (purple shaded areas) and sampled on the altitude
            grid used in the DISORT retrieval algorithm (red line), compared to the constant default profile using ACS-MIR altitudes (dark blue line) or the typical 20 to 100~km assumption (dark blue dotted line), and
            to the linear gradient profile using ACS-MIR altitudes (light blue line).
            \textbf{b.} Vertical profile of the atmospheric extinction coefficient (\kext) at 3.2~\mum{} (red line) derived from
            ACS-MIR observation ORB021039\_2 with the associated uncertainties (red shaded area).
            The light blue areas represent the altitudes where water ice clouds have been identified.}
            \label{fig:example_profile_acs}
        \end{figure}

        Figure~\ref{fig:example_profile_acs} shows the comparison between a cloud profile obtained from ACS-MIR
        observations (ORB021039\_N2) and the typical default for \taucld{} retrievals in the literature, i.e.,
        constant size of 3~\mum{} from 20 to 100~km.
        We can see that if \reff{} is similar at the bottom of the cloud (although ACS-MIR reported a cloud
        starting from 29~km only), it decreases quickly in the observations to reach 0.1~\mum{} at 56~km
        and remain like so up to the top of the cloud at 79~km.
        In this section, we will discuss the impact of using different kinds of \reff{} profiles for the clouds
        in DISORT on the retrieved \taucld:
        \begin{itemize}
            \item Constant profile: \reff{} remains constant regardless of the altitude, this is the assumption
                usually made for such nadir retrievals.
            \item Linear gradient profile: previous studies 
                \citep[e.g.,][]{stcherbinine_2020, stcherbinine_2022e, luginin_2020, liuzzi_2020} have shown that
                the particle size within the clouds typically decreases down to $\sim$~0.1~\mum{} when the
                altitude increases. Thus, we implemented an alternative \reff{} profile where the water ice particle size
                decreases linearly when the altitude increases, ranging from a given value at the bottom altitude
                of the cloud to 0.1~\mum{} at the top.
            \item ACS-MIR profile: here we use the actual \reff{} profile obtained from the correspondent ACS-MIR
                observation, resampled in the altitude grid used in the retrieval algorithm 
                (cf. Figure~\ref{fig:example_profile_acs}).
                As stated above and for simplification purposes, we consider the potential multiple layers
                that may be present in the ACS-MIR data as one single cloud layer, starting from the lowest
                altitude where water ice has been detected, and ending at the higher altitude with water ice.
        \end{itemize}

        \begin{figure}[h]
            \centering
            \includegraphics[width=\columnwidth]{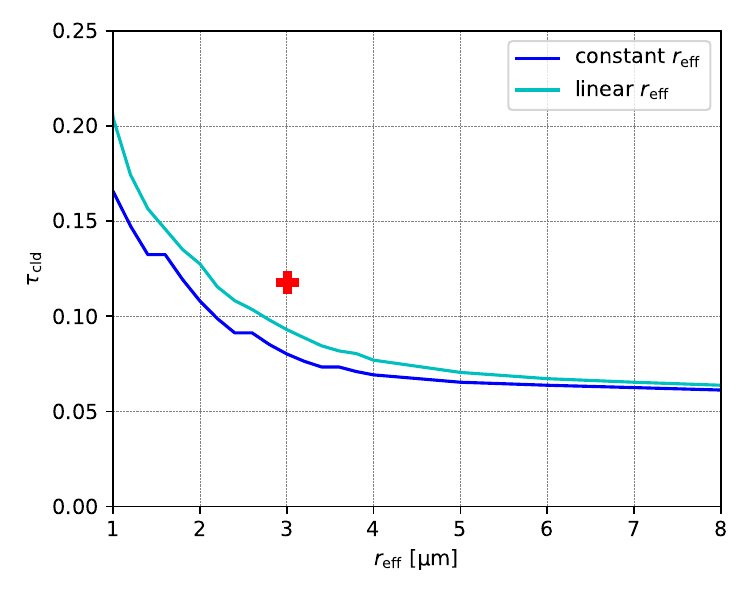}
            \caption{Variation of the retrieved \taucld{} value as a function of the water ice particle size
                within the cloud (EXI observation 20220811T143523) for a cloud with a constant particle size
                (blue line) or using a linear gradient from \reff{} at the bottom of the cloud to
                0.1~\mum{} at the top (cyan line).
                The boundary altitudes of the cloud are taken from the corresponding ACS-MIR observation
                (ORB021039\_N2), i.e., from 29 to 76~km, as well as the maximum altitude for the
                dust, set to 63~km.
                The red cross shows the result obtained using the actual ACS-MIR cloud profile shown
                in Figure~\ref{fig:example_profile_acs} (red line).
                }
            \label{fig:dependence_reff}
        \end{figure}

        \begin{figure}[h]
            \centering
            \includegraphics[width=\columnwidth]{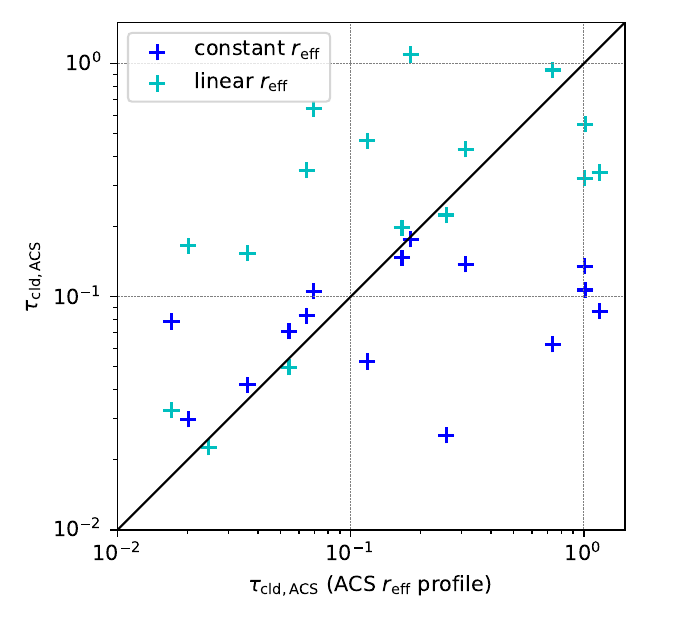}
            \caption{EXI optical depth retrieved using clouds altitudes from ACS-MIR 
                observations and either a constant \reff{} profile of 3~\mum{} (blue) or a linear
                gradient profile ranging from 3~\mum{} at the bottom to 0.1~\mum{} at the top (cyan) as a function of the optical depth retrieved
                using the actual corresponding ACS-MIR cloud profile.
                }
            \label{fig:acs_reff_vs_lin_cst}
        \end{figure}

        Figure~\ref{fig:dependence_reff} shows the \taucld{} values retrieved from the EXI observation
        20220811T143523, using the cloud's altitude from ACS-MIR observation ORB021039\_N2 (from 29 to
        76~km), as a function of the water ice particle size assumed in the retrieval algorithm.
        These computations have been performed for both constant and linear gradient profiles, as well as
        using the actual ACS-MIR \reff{} profile.
        We can see that the retrieved value for \taucld{} depends on the assumed size of the water ice particles
        at the bottom of the cloud (either for the constant or the linear gradient case), especially for
        \reff{} between 1 and 4~\mum. It goes from \taucld~=~0.166 for \reff~=~1~\mum{} to
        \taucld~=~0.069 for \reff~=~4~\mum{} with the constant profile, and respectively from \taucld~=~0.204
        to \taucld~=~0.077 with the linear gradient.
        For larger ice particles, from 4 to 8~\mum{}, it became more stable: \taucld{} goes from 0.069 to 0.061 for the constant profile, and from 0.077 to 0.064 for the linear gradient one.

        We also observe that the difference between the constant and linear gradient profiles is more
        pronounced when only smaller ice particles are present in the cloud: the \taucld{} values retrieved
        with the linear gradient profile are 10~\% to 20~\% larger than the results with the constant
        profile for \reff~$\leq$~4~\mum, but the difference becomes smaller than 5~\% when \reff{} at the
        bottom of the cloud is larger than 6~\mum.

        Smaller water ice particles in the clouds result in more reflective and less absorbing cloud layers
        in the atmosphere, which leads to thicker clouds retrieved in order to match the I/F measured values.
        Thus, introducing smaller particles in the upper part of the cloud with the linear gradient profile
        leads to an increase of the retrieved \taucld{}. However, we can see that the increase of \taucld{}
        due to the transition to a linear gradient profile is of lower magnitude compared to the decrease
        of \reff{} at the bottom of the cloud. This can be explained by the fact that the density of particles
        in the atmosphere is higher at low altitudes, so the lower layers (also filled with more absorbent
        larger water ice particles) will dominate the column-integrated nadir optical depth of the clouds.

        Figure~\ref{fig:dependence_reff} also shows the \taucld{} retrieved using the actual ACS-MIR profile
        as a matter of comparison (red cross). As we can see in Figure~\ref{fig:example_profile_acs}, the
        ACS profile exhibits a decrease of \reff{} from 3~\mum{} at 29~km to 0.1~\mum{} at 56~km, then the
        particle size remains at 0.1~\mum{} up to the top of the cloud at 76~km.
        So, the proportion of very small water ice particles is higher in the ACS-MIR profile compared to the linear gradient
        one.
        Consequently, the value of \taucld{} retrieved with this profile is larger than with the linear
        gradient profile for the same \reff{} (3~\mum): \taucld~=~0.118 for the ACS-MIR profile versus
        \taucld~=~0.093 for the linear gradient one (and \taucld~=~0.080 for the constant profile).
        And so, even though it is not perfect, the use of a linear gradient instead of a constant profile for
        \reff{} in the clouds retrievals improves the retrieved \taucld{} by bringing it more than 40~\% closer
        to what can be obtained using the actual cloud profile.

        The situation is more contrasted on a global view, as can be seen on Figure~\ref{fig:acs_reff_vs_lin_cst}, which shows the \taucld{} values 
        retrieved assuming either a constant \reff{} profile (blue crosses) or a linear gradient
        (cyan crosses) between the boundary altitudes given by ACS-MIR observations, as a function
        of the \taucld{} values retrieved using the actual ACS \reff{} profiles.
        We can see that for the thinner clouds ($\tau_\text{cld} \lesssim 0.15$) the use of the 
        constant profile can provide results closer to the \taucld{} retrieved using the actual
        ACS-MIR profile compared to the linear gradient, but for thicker clouds it is better
        to consider a non-uniform \reff{} profile.
        Actually, we observe in the figure that the uniform profile results in a smaller range of
        \taucld{} (from 0.02 to 0.17) than the linear gradient one (from 0.02 to 1.05).
        For thicker clouds associated with $\tau_\text{cld} \gtrsim 0.15$, even though the 
        \taucld{} values retrieved using the ACS-MIR \reff{} profiles increase (up to 1.2), 
        the \taucld{} values retrieved if assuming a constant profile do not exceed 0.18.

        Thus, neither of the two tested profiles provides results perfectly in line with the
        ones obtained using the actual ACS-MIR \reff{} profiles across the entire range of
        \taucld. The linear gradient will result in lower precision for optically thin clouds,
        but the constant profile will fail to capture the thicker clouds.
        Further studies may try to discuss more advanced vertical profiles of \reff{} such
        as exponential profiles, which may be closer to what has been observed by the
        orbiters \citep[e.g.,][]{stcherbinine_2022e}.

\section{Conclusion}    \label{sec:ccl}
    In this paper, we use the combination of quasi-concomitant observations of the same region of the Martian
    atmosphere by the TGO/ACS-MIR and EMM/EXI instruments to study Martian water ice clouds, and discuss the
    impact of the refinement of the vertical structure of the clouds in radiative transfer models on optical
    depth retrievals performed from nadir imagery.
    The vertical distribution and properties of the aerosols and clouds in the atmosphere are two of the main
    hypotheses that have to be made in order to perform retrievals from nadir observations using radiative
    transfer algorithms. But this information can be retrieved from observations in different geometries, 
    especially in Solar Occultation.
    Thus, we use cloud profiles (altitude and \reff) obtained from ACS-MIR observations to perform optical
    depth (\taucld) retrievals from EXI nadir images at 320~nm, and compare the results to the values obtained
    with a constant cloud profile typically used in this kind of study.

    First, we show that not accounting for the variation of the clouds' altitude with the latitude that can 
    be up to 40~km between the poles and the equator \citep{stcherbinine_2022e} leads to an underestimation
    of \taucld{} up to 25~\% to 50~\% under polar latitudes, and an overestimation of about 10~\% to 25~\% at the equator
    if assuming a bottom altitude of 20~km for all the clouds (depending on the vertical distribution of the dust in the atmosphere).
    
    Then, we show that the accounting for the variation of \reff{} with the altitude in the clouds (with very
    small water ice particles at the top) also significantly affects the retrievals but to a lesser extent than 
    altitude or the size of the ice particles at the bottom of the cloud. The retrieved values of \taucld{} are
    dominated by the size of the larger water ice particles at low altitudes, but the impact of the very small ones
    at the top of the clouds is not negligible.

    Thus, from the results presented in Section~\ref{sec:discussion}, we formulate two recommendations to 
    improve the representation of Martian water ice clouds in radiative transfer models, and thereby the 
    performed optical depth retrievals from nadir observations in future studies:
    \begin{itemize}
        \item Adjust the boundary altitudes of the clouds with latitude and season, from typical ranges that
            can be extracted from recent clouds' climatology, such as presented in \citet{stcherbinine_2022e},
            and adjust the maximal altitude for the dust so it remains below the water ice clouds.
        \item Use of a non-constant \reff{} profile for the clouds. For instance, a profile ranging from
            $\sim$~2-3~\mum{} at the bottom to 0.1~\mum{} at the top would provide a better representation
            of the vertical structure of the clouds and thus more accurate results for the thicker clouds, even if it is not perfect
            either. Future studies may even consider more evolved profiles for the vertical distribution
            of the water ice, with a non-linear decrease of \reff{} that may be closer to what is observed
            by the orbiters \citep[e.g.,][]{stcherbinine_2020, stcherbinine_2022e, liuzzi_2020, luginin_2020, stolzenbach_2023a}.
    \end{itemize}
    
    This study is the first collaboration between the EMM and TGO/ACS science teams and paves the way
    for future synergies between these two active missions currently looking at the Martian atmosphere.
    And beyond that, it highlights the huge potential in developing more synergies between different
    spacecraft and instruments to study the Martian water ice clouds.

\section*{Data availability}    \label{sec:data_availability}
Data from the Emirates Mars Mission (EMM) are freely and publicly available on the EMM Science Data Center (SDC, \url{http://sdc.emiratesmarsmission.ae}).
This location is designated as the primary repository for all data products produced by the EMM team and is designated as long-term repository as required by the UAE Space Agency.
The data available (\url{http://sdc.emiratesmarsmission.ae/data}) include ancillary spacecraft data, instrument telemetry, Level 1 (raw instrument data) to Level 3 (derived science products), quicklook products, and data users guides (\url{https://sdc.emiratesmarsmission.ae/documentation}) to assist in the analysis of the data.
Following the creation of a free login, all EMM data are searchable via parameters such as product file name, solar longitude, acquisition time, sub-spacecraft latitude \& longitude, instrument, data product level, etc.

Data products can be browsed within the SDC via a standardized file system structure that follows the convention:
\texttt{/emm/data/\textless Instrument\textgreater /\textless DataLevel\textgreater /\textless Mode\textgreater /\textless Year\textgreater /\textless Month\textgreater }

Data product filenames follow a standard convention:\linebreak
\texttt{emm\_\textless Instrument\textgreater \_\textless DataLevel\textgreater \textless StartTimeUTC\textgreater \_\textless OrbitNumber\textgreater\linebreak \_\textless Mode\textgreater \_\textless Description\textgreater
\_\textless Kernel-Level\textgreater \_\textless Version\textgreater .\textless FileType\textgreater }

EXI data and users guides are available at: \url{https://sdc.emiratesmarsmission.ae/data/exi}

Raw ACS data are available on the ESA PSA at 
\url{https://archives.esac.esa.int/psa/#!Table%20View/ACS=instrument}.

\section*{Acknowledgments}
This article is dedicated to the memory of Jessica Walsh. For the love of Mars and clouds.

This work was funded by the Emirates Mars Mission project under the Emirates Mars Infrared Spectrometer and the Emirates eXploration Images instruments via The United Arab Emirates Space Agency (UAESA) and the Mohammed bin Rashid Space Centre (MBRSC).
A. S. also acknowledges funding by CNES.

ExoMars is a space mission of ESA and Roscosmos. The ACS experiment is led by IKI Space Research Institute in Moscow. The project acknowledges funding by Roscosmos and CNES. Science operations of ACS are funded by Roscosmos and ESA. Science support in IKI is funded by the Federal Agency of Science Organization (FANO).











\bibliographystyle{cas-model2-names}

\bibliography{bibliography}



\end{document}